\documentclass[submission,creativecommons,sharealike]{eptcs} 
\providecommand{\event}{GandALF 2012}

\title{Automata-based Static Analysis of XML Document Adaptations}
\author{Alessandro Solimando, Giorgio Delzanno, Giovanna Guerrini
\institute{Universit\`a di Genova, Italy
\\
}
\email{\{alessandro.solimando,giorgio.delzanno,giovanna.guerrini\}@unige.it}
}
\def\titlerunning{Automata-based Static Analysis of XML Document Adaptations}
\def\authorrunning{G. Delzanno, G. Guerrini, A. Solimando}

\newcommand{\hide}[1]{}

\usepackage{graphicx}
\usepackage{amssymb,amsmath,amsfonts,amsthm}
\usepackage{subfigure}
\newtheorem{es}{Example}

\begin{document}
\maketitle

\begin{abstract}
The structure of an XML document can be optionally specified by means of XML
Schema, thus enabling the exploitation of  structural information  for efficient
document handling. Upon schema evolution, or when exchanging documents among
different collections exploiting related but not identical schemas, the need
may arise of {\em adapting} a document, known to be valid for a given schema
$S$, to a target schema $S^\prime$. The adaptation may require knowledge of the
element semantics and cannot always be automatically derived. In this paper,  we
present an automata-based method for the static analysis of user-defined XML
document adaptations, expressed as sequences of  XQuery Update update
primitives. The key feature of the method is the use of an automatic inference
method for extracting the type, expressed as a Hedge Automaton, of a sequence
of document updates. The type is computed starting from the original schema $S$
and from rewriting rules that formally define the operational semantics of a
sequence of document updates. Type inclusion can then be used as conformance
test w.r.t. the type extracted from the target schema $S^\prime$. 
\end{abstract}

\section{Introduction}
\label{sec:intro}
XML is a widely employed standard for the representation and exchange of data on
the Web. XML does not define a fixed set of tags, and can thus be used in a
great variety of domains. The structure of an XML document can be optionally
specified by means of a schema, expressed as an XML Schema
\cite{xmlschema_w3c} or as a DTD \cite{dtd_w3c},  and the document structural
information can be exploited for efficient document handling. A given XML schema
can be used by different users to locally store documents valid for the schema.
In a dynamic and heterogeneous world as the Web, updates to such shared schemas
are quite frequent and support for dynamic schema management is crucial to avoid
a diminishment of the role of schemas in contexts characterized by highly
evolving and unstable domains.
As a consequence of a schema update, document validity might need to be
re-established and no automatic way to adapt documents to the new schema may
exist, since the adaptation may require knowledge of the element semantics.
 Moreover, in case of a schema employed in different document
collections, different choices may be taken by individual users handling
different collections, depending on their specific knowledge of the documents in
their collection. Consider for instance the case of an original schema 
containing an optional element {\tt address}. The schema can be updated
by inserting a {\tt zipcode} sibling of {\tt address} (optional sequence), so
that now  either valid documents  do not contain address information at all, or,
if an address is present, the zipcode needs to be present as well. The most
obvious, automatic way to adapt documents could be that of mimic the schema
update thus inserting a zipcode for each address occurrence in a document.
However, in some cases it would be preferable to delete the address instead,
thus restoring the document validity through a different operation (i.e., a
deletion) not directly corresponding to the one occurred on the schema (i.e., an
insertion). 
Moreover, depending on the application contexts, 
only the original schema $S$ and the target schema $S^\prime$ may be known,
while the update sequence that transformed $S$ in $S^\prime$ is not known.
Individual users may thus specify document adaptations, intended to transform
any document valid for $S$ in a document valid for $S^\prime$.
Methods able to validate the document adaptations specified
by individual users are then useful to avoid
the expensive run-time revalidation of documents resulting from the application
of such adaptations.

In this paper, we present an automata-based method, called HASA (Hedge Automata
Static Analyzer), for the static analysis of XML document adaptations, expressed
as sequences of XQuery Update (XQUF) \cite{xqueryupdatefacility_w3c} update
primitives. 
The key feature of HASA is the use of an automatic inference method for
extracting the type of a sequence of document updates. The type is computed
starting from a static type assigned to an XML schema and from rewriting rules
that formally define the operational semantics of a sequence of document
updates. Type inclusion can then be used as conformance test w.r.t. the type
extracted from the updated XML schema.
Our types are represented via Hedge Automata (HA). 
Hedge Automata are a very flexible and general tool for manipulating trees.  
Indeed they can handle ranked and unranked ordered trees. 
Furthermore, validation algorithms for XML schemas are naturally expressed via
Hedge Automata.
It comes natural to extract the type of an XML schema  in form of an Hedge
Automaton~\cite{TaxXML}.
We exploit this feature in order to define a HA2HA transformation that
produces the type of a document adaptation.
Specifically, HASA takes as input two XML schemas $S$ and $S^\prime$ such that
$S^\prime$ is an evolution of (i.e., the result of a, possibly unknown, sequence
$U$ of updates on) $S$. For each schema, we automatically generate the
corresponding types in form of the Hedge Automata $A$ and $A^\prime$. The user
now provides a sequence of document updates $u_1,...,u_k$  (document adaptation)
to make instances of $S$ conform to the new schema $S^\prime$. Given $A$, we
compute the Hedge Automaton $A_1=Post(u_1,A)$ that recognizes the documents in
$A$ after the modification $ u_1$. We then repeat the computation for $
u_2,...,u_k$ producing a Hedge Automaton $ A_k$ that recognizes the documents
after the complete sequence of updates. The resulting automaton $ A_k$ can now
be compared with the Hedge Automaton $A^\prime$. If the language of $A_k$ is
included in that of $A^\prime$, the proposed document adaptation surely
transforms a document known to be valid for $S$ in a document valid for
$S^\prime$.  
If inclusion does not hold, we use the automaton $A_k$ as a tester to identify
documents that do not conform to $S^\prime$ (i.e., testing whether the execution
of the automaton $A_k$ over the document corresponds to an accepting
computation).

In this paper we focus our attention on the technical details underlying the
design of the HASA module.
Specifically, our technical contribution is as follows:
First, we introduce a parallel rewriting semantics for modelling the effect of a document update
	 on a term-based representation of 	XML documents. 
	Our semantics is based on a representation of document updates as special    
      types of term rewriting systems 
	\cite{jacqemard_rusinowitch}, and on a parallel semantics for modeling
the simultaneous application of a rewrite rule to each node that satisfies its
enabling conditions 
	(we consider here node selection only).
	As an example, we model renaming of label $a$ into label $b$ as a
rewrite rule $r=a(x)\rightarrow b(x)$ where  
      $x$ is a variable that denotes 
	an arbitrary list of subtrees. A document is represented as a tree $t$.
 	Renaming must be applied to all occurrences of label $a$ in $t$, i.e., as a maximal parallel rewriting step
 	computed w.r.t. $r$.
	A parallel rewriting semantics needs to be considered,  instead of the more standard sequential semantics 
      used in rewriting systems, to capture the semantics of more complex operations like document insertion.
   In case of document insertions, indeed, a sequential semantics
	may lead to incorrect rewriting steps (e.g., to recursively modify a subtree being inserted). 
	
We then move to the symbolic computation of types, i.e., of Hedge
Automata that represent the effect of applying a document adaptation on the
initial automaton $A$. More specifically, we give HA2HA transformations that
simulate the effect of a parallel application of each type of update rules. 
A symbolic algorithm is defined to compute $Post$ as a Hedge Automata transformation and proved correct w.r.t. our parallel rewrite semantics. This is the core operation of our HASA approach. 
  Differently from other automata-based transformation approaches
\cite{Touili_computingtransitive}, we are interested here in calculating the effect of a 
  single document update and not of its transitive closure.
  
Finally, a proof of concept implementation of the HASA module has been developed as a
modification of the LETHAL library.

The paper is organized as follows.
In Section \ref{sec:prelim} some preliminary notions are introduced.
Section \ref{subsec:hedgeautomata} introduces Hedge Automata as a formalism to describe
XML schemas, while Section \ref{sec:parallel}  is devoted to XQuery
Update primitives and to the 
corresponding update rewrite rules, with their parallel rewriting semantics.
  Section \ref{sec:hasa} describes the symbolic algorithm underlying the HASA module.
  Section \ref{sec:conclusionrw} concludes by discussing related work and future
research directions.

\section{Preliminaries}
\label{sec:prelim}
In this section we introduce the notations and definitions (mainly from
\cite{tata2007}) used in the remainder of the work.
We refer to \emph{terms} and
\emph{trees} as synonyms as in \cite{tata2007}.
Given a string $s \in L \subseteq \Sigma^*$ the set of its prefixes w.r.t. $L$
is defined as $Pref_L(s) = \{ t \mid s = tu \wedge t,u \in L\}$.
When the language is clear from the context we use $Pref$ instead of $Pref_L$.
Given a language $L \subseteq \Sigma^*$ we call {prefix language} the set of the
prefixes of the elements of $L$: $Prefixes(L) = \bigcup_{s \in L} Pref_L(s)$. A
language $L \subseteq \Sigma^*$ is said  {prefix-closed} if $Prefixes(L) = L$,
that is, if the language contains every possible prefix of every string
belonging to the language itself.

A {term} is an element of a ranked alphabet defined as ($\Sigma$,
\textit{Arity}), where $\Sigma$ is a finite and nonempty alphabet,
\textit{Arity}$\colon \Sigma \rightarrow \mathbb{N}$ is a function that
associates a natural number, called {arity} of the symbol, with every
element of $\Sigma$. The set of symbols with arity $p$ is denoted as $\Sigma_p$
(for the sake of conciseness we will use a compact notation, e.g., $f(,,)$ is a
term contained in $\Sigma_3$). $\Sigma_0$ is called the set of
{constants}. Let $\mathcal{X}$ be a set of {variables}, disjoint
from $\Sigma_0$. The set $T(\Sigma, \mathcal{X})$ of the terms over $\Sigma$ and
$\mathcal{X}$ is defined as: (1) $\Sigma_0 \subseteq T(\Sigma, \mathcal{X})$,
(2) $\mathcal{X} \subseteq T(\Sigma, \mathcal{X})$, (3) if $f \in \Sigma_p$, $p
> 0$ and $t_1, \ldots, t_p \in T(\Sigma, \mathcal{X})$, then $f(t_1, \ldots,
t_p) \in T(\Sigma, \mathcal{X})$. If $\mathcal{X} = \emptyset$ we use
$T(\Sigma)$ for $T(\Sigma, \mathcal{X})$ and its elements are called
{ground terms}, terms without variables. {Linear terms} are the
elements of $T(\Sigma, \mathcal{X})$ in which each variable occurs at most
once.

A finite and ordered {ranked tree} $t$ over $\Sigma$ is a map from a
set $\mathcal{P}os(t) \subseteq \mathbb{N}^*$ into a set of labels $\Sigma$,
with $\mathcal{P}os(t)$ having the following properties: (1) $\mathcal{P}os(t)$
is finite, nonempty and prefix-closed, (2) $\forall p \in \mathcal{P}os(t)$, if
$t(p) \in \Sigma_n$ and $n > 0$, then $\{j \mid p.j \in \mathcal{P}os(t)\} =
\{1,\ldots,n\}$, (3) $\forall p \in \mathcal{P}os(t)$, if $t(p) \in \Sigma_0
\cup \mathcal{X}$, then $\{j \mid p.j \in \mathcal{P}os(t)\} = \emptyset$.
\emph{Root}$(t) = t(\epsilon)$ is called {root} of the tree.
An {unranked tree} $t$ with labels belonging to a set of
unranked symbols $\Sigma$ is a map $t \colon \mathbb{N}^* \rightarrow \Sigma$ with a
domain, denoted as $\mathcal{P}os(t)$, with the followings properties: (1)
$\mathcal{P}os(t)$ is a finite, nonempty and prefix-closed, (2) for every $p
\in \mathcal{P}os(t)$  $\{ j \mid p.j \in \mathcal{P}os(t)\} =
\{1,\ldots,k\}$ for some $k \geq 0$. The set of unranked trees over $\Sigma$ is
denoted as $T(\Sigma)$.
The {subtree} $t|_p \in T(\Sigma, \mathcal{X})$ is the
subtree in position $p$ in a tree $t \in T(\Sigma, \mathcal{X})$ such
that $\mathcal{P}os(t|_p) = \{ j \mid p.j \in  \mathcal{P}os(t)\}$ and $\forall
q \in \mathcal{P}os(t|_p)$ . $t|_p(q) = t(p.q)$.

An example of unranked tree is $t = a(b(a, c(b)), c, a(a,c))$. Note that the
same label can be used in different nodes which may have a different number of
children (an arbitrary but finite value). An example of subtree is $t|_1
= b(a, c(b))$.
%

\section{Hedge Automata (HA) and XML Documents}
\label{subsec:hedgeautomata}
Tree Automata (TA) are a natural generalization of finite-state automata to
define languages over ranked finite trees (instead of finite words). TA can
naturally be used as a formal support for document validation~\cite{murata_schema_trans,murata:dtd_transf}.
In this setting, however,  it is often more convenient to consider more general classes of
automata, like Hedge and Sheaves Automata, to manipulate both ranked and
unranked trees. Indeed, in XML documents the number of children of a node with a certain label is
not fixed a priori, and different nodes sharing the same label may have a
different number of children. Hedge Automata (HA) are a suitable formal tool for reasoning on a
representation of XML documents via unranked trees. 
HA are a generalization of TA because in the latter only ranked symbols are
supported and the horizontal languages are fixed sequences of states whose
length is the rank of the considered symbol.
We introduce  the main
ideas underlying HA definition in what follows.

Given an unranked tree $a(t_1, \ldots, t_n)$ where $n \ge 0$, 
the sequence $t_1, \ldots, t_n$ is called {hedge}. For $n = 0$ we have an
empty sequence, represented by the symbol $\epsilon$. The set of hedges
over $\Sigma$ is $H(\Sigma)$. Hedges over $\Sigma$ are inductively defined in
\cite{murata-Bruggemann-Klein01regulartree} as follows: the empty sequence
$\epsilon$ is a hedge, if $g$ is a hedge and $a \in \Sigma$, then $a(g)$ is a
hedge, if $g$ and $h$ are hedges, then $gh$ is a hedge.
For instance, given a tree $t=a(b(a,c(b)), c, a(a,c))$, the corresponding hedges having as
root nodes the children of $Root(t)$ are $b(a,c(b))$, $c$ and $a(a,c)$.

A {Nondeterministic Finite Hedge Automaton} (NFHA) defined over $\Sigma$
is a tuple $M = (Q, \Sigma, Q_f, \Delta)$ where $\Sigma$ is a finite and non
empty alphabet, $Q$ is a finite set of states, $Q_f \subseteq Q$ is the set of
final states, also called accepting states, $\Delta$ is a finite set of
transition rules of the form $a(R) \rightarrow q$, where $a \in \Sigma$, $q \in Q$ and $R
\subseteq Q^*$ is a regular language over $Q$.
Regular languages denoted as $R$ that appear in rules belonging to $\Delta$
are said {horizontal languages}, represented with Nondeterministic Finite
Automata (NFA).
The use of regular languages allows us to consider unranked trees.
For instance, $a(q^*)$ matches a node $a$ with any number of subtrees generated
by state $q$.

A {computation} of $M$ over a tree $t \in T(\Sigma)$ is a tree $M||t$
having the same domain of $t$ and for which, for every element $p \in
\mathcal{P}os(M||t)$ such that $t(p) = a$ and $M||t(p) = q$, a rule $a(R)
\rightarrow q$ in $\Delta$ must exist such that, if $p$ has $n$ successors
$p.1,\ldots,p.n$ such that $M||t(p.1) = q_1, \ldots, M||t(p.n) = q_n$, then $q_1
\cdots q_n \in R$. If $n = 0$ (that is, considering a leaf node) the empty
string $\epsilon$ must belong to the language $R$ of the rule to be applied to
the leaf node.
A tree $t$ is said to be {accepted} if a computation exists in which the root
node has a label $q \in Q_f$. The {accepted language} for an automaton $M$,
denoted as $L(M) \subseteq T(\Sigma)$, is the set of all the trees accepted by
$M$.
\begin{figure}[htp!]
\centering
\subfigure[Tree $t$ representing a true Boolean formula.]{%
  \label{fig:es_nfha}
  \includegraphics[width=0.3\textwidth]{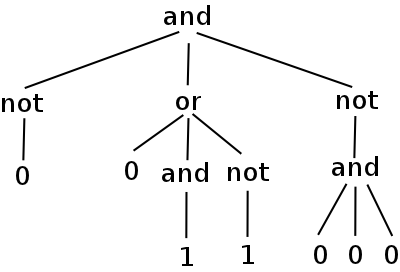}
}
\subfigure[Accepting computation of the automaton $M$ over tree $t$.]{%
  \label{fig:es_comp:nfha}
  \includegraphics[scale=0.1,
width=0.3\textwidth]{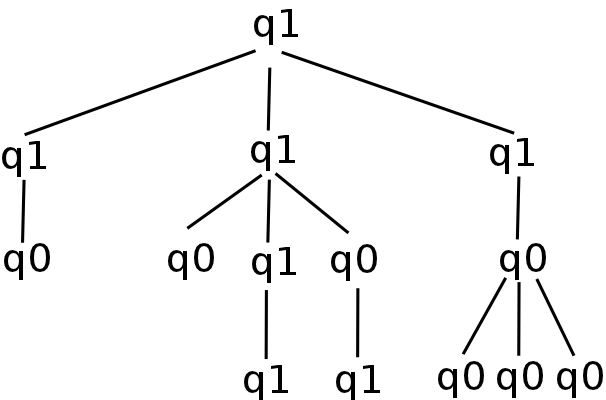}
}%
\caption{An example of tree $t$ (left) and the computation $M||t$ of the
automaton $M$ over $t$ (right).}
\end{figure}
\\
As an example, consider the NFHA $M = (Q, \Sigma, Q_f, \Delta)$ where $Q =
\{q_0, q_1\}$, $\Sigma = \{0, 1, not, and, or\}$, $Q_f = \{q_1\}$ and $\Delta =
\{not(q_0) \rightarrow q_1, not(q_1) \rightarrow q_0, 1(\epsilon) \rightarrow
q_1, 0(\epsilon) \rightarrow q_0, and(Q^*q_0Q^*) \rightarrow q_0, and(q_1q_1^*)
\rightarrow q_1, or(Q^*q_1Q^*) \rightarrow q_1, or(q_0q_0^*) \rightarrow
q_0\}$. Figure \ref{fig:es_nfha} shows a tree $t$ representing a
Boolean formula. Figure \ref{fig:es_comp:nfha} shows the accepting
computation of the automaton $M$  (i.e., $M||t(\epsilon) = q_1 \in Q_f$).
Note that though $and, or$ are binary logic operators, we used their associativity to treat
them as unranked symbols. The equivalent TA differs from the HA only in
the rules for these binary operators $\Delta = \{\ldots, and(q_0, q_0)
\rightarrow q_0, and(q_0, q_1) \rightarrow q_0, and(q_1, q_0) \rightarrow q_0,
and(q_1, q_1) \rightarrow q_1, or(q_0, q_0) \rightarrow q_0, or(q_0, q_1)
\rightarrow q_1, or(q_1, q_0) \rightarrow q_1, or(q_1, q_1) \rightarrow q_1,
\ldots \}$. 

A NFHA $M = (Q, \Sigma, Q_f, \Delta)$ is said {normalized} if, for each
$a \in \Sigma, q \in Q$ at most one rule $a(R) \rightarrow q \in \Delta$ exists.
Since string regular languages are closed under union \cite{tata2007}, it is always
possible to define a normalized automaton starting from a non normalized NFH.
Every pair of rules $a(R_1) \rightarrow q$ and $a(R_2) \rightarrow q$ belonging
to $\Delta$ is substituted by the equivalent rule $a(R_1 \cup R_2) \rightarrow
q$.

Given two NFHA $M_1$ and $M_2$, the {\em inclusion test} consists in checking 
whether $L(M_1) \subseteq L(M_2)$. It can be reduced to the emptiness 
test for HA ($L(M_1) \subseteq L(M_2) \Leftrightarrow L(M_1) \cap (T(\Sigma)
\setminus L(M_2)) = \emptyset$). 
Inclusion test is decidable, since complement, intersection and emptiness of HA
can be  algorithmically executed~\cite{tata2007}.

\section{XQuery Update Facility as Parallel Rewriting} 
\label{sec:parallel}
\begin{table}[t]
\centering
\begin{tabular}{|c|c|}
\hline
update rule& XQUF primitive update operation\\
\hline
$a(x) \rightarrow b(x)$ & $REN$\\
$a(x) \rightarrow p$ & $RPL$\\
$a(x) \rightarrow ()$ & $DEL$\\
$a(x) \rightarrow a(px)$ & $INS_{first}$\\
$a(x) \rightarrow a(xp)$ & $INS_{last}$\\
$a(xy) \rightarrow a(xpy)$ & $INS_{into}$\\
$a(x) \rightarrow pa(x)$ & $INS_{before}$\\
$a(x) \rightarrow a(x)p$ & $INS_{after}$\\
\hline
\end{tabular}
\renewcommand{\tablename}{Table}
\caption{XQUF primitives. $a$ and $b$ are XML tags, $p$ is a state
of an HA, and $x,y$ are free variables that denote arbitrary sequences of
trees.}
\label{tab:ufo}
\end{table}
XQUF \cite{xqueryupdatefacility_w3c} is an update language for XML. Its
expressions are converted into an intermediate format called Pending Update List
(PUL). In this paper we consider a formulation of PULs as a special class of
rewriting rules defined on term symbols and types (states of Hedge Automata) as
suggested in \cite{jacqemard_rusinowitch}. More specifically, we use the set of
rewriting rules defined in Table~\ref{tab:ufo}. The idea is as follows. Target
node selection is based on the node label only  (and not on hierarchical
relationships among nodes). In Table~\ref{tab:ufo}, $a$ and $b$ are node labels,
and $p$ is an automaton state that we interpret as type declaration (it defines
any tree accepted by state $p$). The supported update primitives 
allows for
renaming an element ($REN$), replacing an element and its content ($RPL$),
deleting an element ($DEL$), inserting a subtree as a first, last, or an
arbitrarily positioned child of an element ($INS_{first}, INS_{last},
INS_{into}$, respectively) and inserting a subtree before or after a given
element ($INS_{before}, INS_{after}$, respectively).
According to \cite{xqueryupdatefacility_w3c}, the semantics (i.e., the actual insert
position) of $INS_{into}$ is implementation dependent.
In real systems, in several cases the operation is simply not provided or 
it is implemented either as $INS_{first}$ or as $INS_{last}$.

To illustrate the update rules, consider for instance the rule $REN$ $a(x)\rightarrow b(x)$.
Given a tree $t$,  the rule must be applied to every elements with label $a$.
Indeed, $x$ is a free variable that matches any sequence of subtrees.
If the rule is applied to element $e$ with label $a$ and children $t_1,\ldots,t_k$, 
the result of its application is the renaming of $a$ into $b$, 
i.e., the subterm $a(t_1,\ldots,t_k)$ is replaced by the subterm $b(t_1,\ldots,t_k)$.
Consider now the rule $INS_{first}$ defined as $a(x)\rightarrow a(px)$, 
where $p$ is a type (a state of an HA automaton).
Given a tree $t$,  the rule must be applied to every element with label $a$.
If the rule is applied to element $e$ with label $a$ and children $t_1,\ldots,t_k$, 
the result of its application is the insertion of a (nondeterministically
chosen) term $t$ of type $p$ to the left of the current set of children, i.e.,
the subtree $a(t_1,\ldots,t_k)$ is replaced by the subterm
$a(t,t_1,\ldots,t_k)$. To model the application of an XQUF primitive rule of
Table~\ref{tab:ufo} to each occurrence in a term, we define next a maximal
parallel rewriting semantics denoted via the relation $\Rightarrow_r$ (formally
defined in \cite{techrep}). In the previous example, $INS_{first}$
inserts a tree of type $p$ to the left of the children of each one of the
$a$-nodes in the term $t$.

To assign a formal meaning to our rewriting system, we first define the general
class of rules we adopt here and then we specify the semantics needed to
model document adaptations.
\subsection{Parameterized Hedge Rewriting System}
Let $A = (\Sigma, Q, Q_f, \Delta)$ be an HA (whose states are used as
types in the rules). A Parameterized Hedge Rewriting System (PHRS)
\cite{jacqemard_rusinowitch} $R/A$ is a set of hedge rewriting rules of the
form $L \rightarrow R$, where $L \in H(\Sigma, \mathcal{X})$, and $R \in
H(\Sigma \uplus Q, \mathcal{X})$.
As in Table~\ref{tab:ufo}, we restrict our attention to linear rewriting rules
(with a single occurrence of each variable in the left-hand and right-hand
side). In \cite{Touili_computingtransitive} and \cite{jacqemard_rusinowitch} the
operational semantics of update rules is sequential because it applies a single
rewriting rule at each step (both the rule and the term to which it is applied
are chosen in a nondeterministic way). An XML document update, instead, has a
global effect. For instance, when renaming a label in an XML schema, all the
nodes having that label must be renamed. Such an update may be expressible
through maximal steps of sequential applications of the $REN$ rewriting rule.

Maximal sequential rewrite is not applicable to insertion rules like
$INS_{first}= a(x) \rightarrow a(px)$: sequential applications of $INS_{first}$
may select a single target node more than once, thus yielding incorrect results.
For instance, let 
$t = a(a(b,c), b)$ 
be the tree representation of an XML document and $t^\prime = d(e)$ the tree
corresponding to an XML fragment. Consider the insertion of $t^\prime$ into $t$
as first child of all the nodes labelled by $a$ through the operation $r$
defined as $a(x) \rightarrow a(t^\prime x)$. If we use the standard sequential
semantics of term rewriting we need two applications of rule $r$, one for each
node matching the left-hand side. This leads to terms like 
$t_1=a(a(d(e), d(e),b, c), b)$, $t_2=a(d(e), d(e), a(b, c), b)$,
and $t_3=a(d(e), a(d(e), b,c), b)$.
The intended semantics of $INS_{first}$ requires $r$ to be applied to all
matching occurrences of $a(x)$ in $t$, therefore only the latter term corresponds 
to a correct transformation of $t$.
\subsection{Parallel Rewriting}
In order to capture the meaning of update rules as document adaptation we
introduce a new parallel rewriting semantics for PHRS. In what follows we give the main
ideas underlying the formal definition which is presented in \cite{techrep}.

Given a term $t$ and an update rule $r = L \rightarrow R$, we first identify the
set of positions in the term $t$ that match the left-hand side $L$ of the rule. 
The set of positions in $t$ (strings of natural numbers, see preliminaries) is
ordered according to the lexicographic ordering $<_{lex}$. $t|_{p}$ denotes the
subtree at position $p$.
Let $Target(t,r)$ be the $<_{lex}$-ordered list of nodes that match the
left-hand side of rule $r$.
A substitution is a map $\{x_1 \leftarrow t_1 , \ldots , x_n \leftarrow t_n\}$ that substitutes 
$x_i$ with $t_i$, where $i \in [1,n]$.  
A substitution is extended to terms with variables in the natural way.

A parallel rewriting step of a rule $r$ on a tree $t$ 
is defined as a transformation of $t$ into a new term $t^\prime$ obtained
as follows. 
The tree $t$ is visited bottom-up starting from its leaves.
Every time a node $a(t)$ that matches the rule $a(x)\rightarrow R$ 
via the substitution $\sigma$ is encountered, we replace $a(t)$ with 
$R\sigma$ and then we move to the parent of the node.

The transformation is defined following a descreasing lexicographic ordering in
$Target(t,r)$. To process the current position, we first compute the contexts in
which the rewrite step takes place (to preserve the part of the tree that is not
rewritten), and then we replace the matched left-hand side with $R\sigma$.
The $INS_{into}$ rule requires some care because the insertion position is
nondeterministically selected among the set of children of the matched node.

Now we show an example that involves the $INS_{after}$ rule, for the term 
$t = b(c,d(c(a),a))$, and the rule $r = c(x) \rightarrow c(x)p$ where $p$ is a 
type that contains at least the terms $t_2=a(b)$ and $t_1=a(c(a),c(a))$
as possible instances.
The set of positions in $t$ is defined as  $\mathcal{P}os(t) = \{\epsilon, 1, 2,
2.1, 2.2, 2.1.1\}$. The rule $r$ matches nodes of $t$ at positions  $1$ and
$2.1$.
\begin{itemize}
\item 
We start from the greatest position $2.1$ and compute the context 
$C_2$ defined  by the term $b(c,d(y,a))$ 
(a context is obtained by replacing the subtree at position $2.1$ with a fresh
variable $y$). The substitution $\sigma_2 = \{x \leftarrow (a)\}$ is the result
of matching $c(x)$ with $c(a)$. We can now rewrite the context $C_2[y]$ as
$C_2[R_2\sigma_2]$ where $R_2$ is obtained by instantiating $p$ with term $t_2$.
This gives us the intermediate tree $b(c,d(c(a),\underline{a(b)},a))$ (the new
subtree is underlined).
\item 
We now consider the position $1$, extract the context $C_1 = b(y,d(c(a),a(b),a))$ and consider the matching 
substitution $\sigma_1 = \{x \leftarrow \epsilon\}$ between $c(x)$ and $c$. 
We apply the rewrite step by substituting $y$ with $R\sigma_1$ 
and obtain the new term $C_1[R_1\sigma_1]=b(c,\underline{a(c(a),c(a))},a(b),d(c(a),a(b),a))$ 
(the inserted subtree is underlined) that corresponds to the result of the parallel rewriting step.
\end{itemize}
We remark that a rule with a type term like $p$ may yield different instantiations of $p$ 
in the same parallel step (as in the previous example).
The definition can be extended in a natural way to a set $R$ of update rules. We use 
$\Rightarrow_R$ to denote the resulting relation and $\Rightarrow^*_R$ to denote its 
transitive-reflexive closure.

Finally, we define $Post_{R/A}(S)$, where $S\subseteq T(\Sigma, \mathcal{X})$
and $R/A$ is a PHRS based on update rules, as the language obtained by
a single application of rules in the set $R$ to each element of $S$ through the
parallel rewriting semantic associated to update rules. When  $R$ and $A$ are clear from the
context the shorthand $Post(S)$ is employed.

\section{Hedge Automata-based Static Analysis (HASA)}
\label{sec:hasa}
In this section we describe the symbolic algorithm underlying the HASA module.
As mentioned in the introduction, our goal is to effectively compute the effect
of a document adaptation on each tree that is accepted by a given HA $A$.
For this purpose, fixed an update rule $r$ we define a HA transformation from $A$ to a 
new HA $A^\prime$ such that 
$L(A^\prime)=\{t^\prime \mid t\Rightarrow_r t^\prime,\ t\in L(A)\}$.
In order to define such a transformation we need to carefully operate on the
vertical component of $A$ (rewriting rules that accept the node labels) as well as
on the horizontal languages (e.g., for operations like insertion). The
$INS_{into}$ rule is discussed at the end of the section. We anticipate that the
nondeterminism in the choice of the insertion position may introduce the need of
considering several alternatives for the $Post$ computation for the same
instance rule. In practical implementations this can be avoided since the semantics in 
$INS_{into}$ rule is always resolved in favour of some fixed insertion position. 
In what follows we provide some examples of the construction. 
The correctness proof of the algorithm w.r.t. to our parallel semantics is 
given in \cite{techrep}, due to space limitations.

%
%
Given two HA $A = (\Sigma_o, P, P_f, \Theta)$ (the HA that describes the types
occurring in the update rules) and $A_L = (\Sigma_L, Q_L, Q^f_L, \Delta_L)$ (the
HA that describes the structure of a set of documents) such that $A$ and $A_L$
are normalized automata, $P \cap Q_L = \emptyset$, $L = L(A_L)$, we define the HA 
$A^\prime = (\Sigma := \Sigma_o \cup \Sigma_L, P \cup Q_L, Q^f_L,\Delta^\prime)$
 such that $L(A^\prime) = Post_{R/A}(L)$. 
The transition relation $\Delta^\prime$ is defined on top of individual laws,
one for each type of update rewriting rule.

For each $a \in \Sigma$, $q \in Q_L$, we denote with $L_{a,q}$ the horizontal language of the
unique rule $a(L_{a,q}) \rightarrow q \in \Delta_L$, accepted by the NFA 
$B_{a,q} = (Q_L, S_{a,q}, i_{a,q}, \{f_{a,q}\}, \Gamma_{a,q})$.

As a preliminary operation we need to expand the alphabet of each automaton that
recognizes the horizontal languages, from $Q_L$ to $P \cup Q_L$. For each of the
following rules we assume $p \in P$, which allows only hedges included
in the language $L(A)$ to be inserted.

For the operations $INS_{before}$, $INS_{after}$, $RPL$ and $DEL$ either 
some states $q \in Q_L$ involved in a change could be shared among different symbols in
$\Sigma$, or two rules $a(L_a) \rightarrow q, b(L_b) \rightarrow
q \in \Delta_L$ could exist such that $a \neq b$. To avoid an unwanted change for 
symbol $b$ a fresh state $q^{fresh}_a \notin P \cup Q_L$ is created and, for each
rule in which the label $a$ and the state $q$ appear simultaneously,  a
copy of this state is created and $q$ is replaced by $q^{fresh}_a$. As last step,
$q^{fresh}_a$ is added to $Q_L$ and to any other alphabet belonging to the horizontal
languages, while updating also their transitions. These changes must be applied
before any other modification.

In the following we present the modification rules for each XQUF primitive rule.
\subsubsection*{Renaming: $REN$}
Given the rule $a(x) \rightarrow b(x) \in R/A$, where $a, b \in
\Sigma$, for each $q \in Q_L$ such that $L(B_{a,q}) \neq \emptyset$ holds, then 
if $L(B_{b,q})=\emptyset$ we define $B_{b,q} := B_{a,q}$, by changing the indexes of the various elements. 
By contrast, if $L(B_{b,q})\neq\emptyset$, we define a new version of
$B_{b,q}$ as the automata that recognize the union of $L(B_{a,q})$ and $L(B_{b,q})$, 
i.e.,
$B_{b,q}=(Q_L,S_{a,q} \uplus S_{b,q} \uplus i_{ab,q}, i_{ab,q}, \{f_{a,q}\} \uplus
\{f_{b,q}\}, \Gamma_{a,q} \uplus \Gamma_{b,q} \uplus \{(i_{ab,q}, \epsilon,
i_{a,q}), (i_{ab,q}, \epsilon, i_{b,q})\})$.
Finally, we remove the rules of the form $a(L_{a,q}) \rightarrow q$ from
$\Delta_L$, where $q \in Q_L$ and we add the corresponding rule $b(L_{b,q})
\rightarrow q$ for each deleted transition. These changes on one hand allow the
automaton to accept the label $b$ where the old automaton accepts label $a$. On
the other hand they preserve the ``behaviour'' of the label $b$ in the horizontal
languages it can be evaluated.

\subsubsection*{Insert first: $INS_{first}$}
The rule $a(x) \rightarrow a(px) \in R/A$ leads to change
the automaton $B_{a,q}$, for each $q \in Q_L$ such that $L_{a,q} \neq
\emptyset$. A fresh state $q^{fresh}_{a,q}$ such that $q^{fresh}_{a,q} \notin
S_{a,q}$ is created, then it is added to $S_{a,q}$ and used as an initial state.
After that, if $\Gamma_{a,q} = \emptyset$ holds, the transition $(q^{fresh}_{a,q}, p,
f_{a,q})$ is added to $\Gamma_{a,q}$. Otherwise, for each transition of the form
$(i_{a,q}, y, q_y) \in \Gamma_{a,q}$, where $i_{a,q}$ is an initial state, 
$y \in P \cup Q_L$, $q_y \in S_{a,q}$, a transition of the form $(q^{fresh}_{a,q}, p, i_{a,q})$ is added to
$\Gamma_{a,q}$.

\begin{figure}[htp]
\centering
\includegraphics[width=0.75\textwidth]{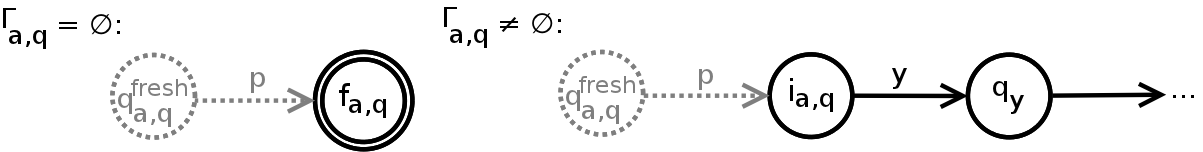}
\renewcommand{\figurename}{Figure}
\caption{The changes to the horizontal automaton due to rule $INS_{first}$ are
depicted as grey texts and dotted lines.}
\label{fig:INSfirst_cap7.png}
\end{figure}

\subsubsection*{Insert last: $INS_{last}$} 
The rule $a(x) \rightarrow a(xp) \in R/A$ leads to change
the automaton $B_{a,q}$, for each $q \in Q_L$ such that $L_{a,q} \neq
\emptyset$. A fresh state $q^{fresh}_{a,q}$ such that $q^{fresh}_{a,q} \notin
S_{a,q}$ is created, added to $S_{a,q}$ and used as 
final state.
Then, if $\Gamma_{a,q} = \emptyset$ holds, the transition $(i_{a,q}, p,
q^{fresh}_{a,q})$ is added to $\Gamma_{a,q}$. Otherwise, for each rule of the
form $(q_y, y, f_{a,q}) \in \Gamma_{a,q}$, where $y \in P \cup Q_L$, $q_y \in
S_{a,q}$,  a transition of the form $(f_{a,q}, p, q^{fresh}_{a,q})$ is added to
$\Gamma_{a,q}$.

\begin{figure}[htp]
\centering
\includegraphics[width=0.75\textwidth]{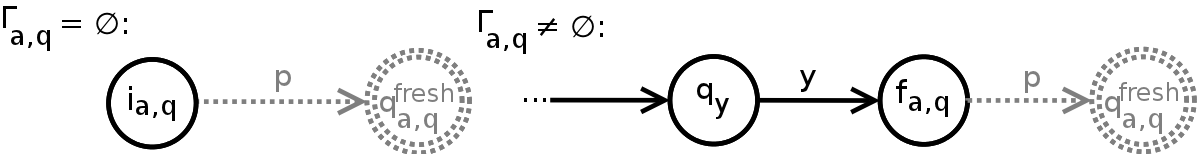}
\renewcommand{\figurename}{Figure}
\caption{The changes to the horizontal automaton due to rule $INS_{last}$ are
depicted as grey texts and dotted lines.}
\label{fig:INSlast_cap7.png}
\end{figure}

\subsubsection*{Insert before: $INS_{before}$}
For the rule $a(x) \rightarrow pa(x) \in R/A$ we need to
modify each horizontal language in which a state $q \in Q_L$ such that
$L(B_{a,q}) \neq \emptyset$ may occur. For each $q \in Q_L$ such that
$L(B_{a,q}) \neq \emptyset$  a fresh state $q^{fresh}_{a,q}$ is created such that
$q^{fresh}_{a,q} \notin S_{b,z}$, for each $b \in \Sigma$ and $z \in Q_L$. Then,
$q^{fresh}_{a,q}$ is added to $S_{b,z}$ if at least one transition of the form
$(s, q, s^\prime) \in \Gamma_{b,z}$, where $s, s^\prime \in S_{b,z}$ exists.
These transitions are changed to $(s, p, q^{fresh}_{a,q})$, after that the
corresponding transitions to $(q^{fresh}_{a,q}, q, s^\prime)$ are added to
$\Gamma_{b,z}$.

\begin{figure}[htp]
\centering
\includegraphics[width=0.6\textwidth]{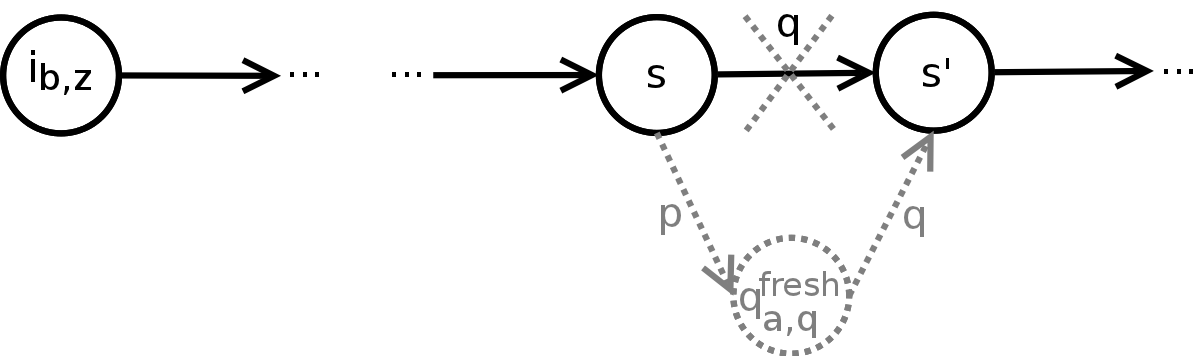}
\renewcommand{\figurename}{Figure}
\caption{The changes to the horizontal automaton due to rule $INS_{before}$ are
depicted as grey texts and dotted lines.}
\label{fig:INSbefore_cap7.png}
\end{figure}

\subsubsection*{Insert after: $INS_{after}$}
For the rule $a(x) \rightarrow a(x)p  \in R/A$ we need to
modify each horizontal language in which a state $q \in Q_L$ such that
$L(B_{a,q}) \neq \emptyset$ may occur. For every $q \in Q_L$ such that
$L(B_{a,q}) \neq \emptyset$ a fresh state $q^{fresh}_{a,q}$ is created such that
$q^{fresh}_{a,q} \notin S_{b,z}$, for each $b \in \Sigma$ and $z \in Q_L$.
This new state is added to $S_{b,z}$ if at least one transition of the form $(s,
q, s^\prime) \in \Gamma_{b,z}$, where $s, s^\prime \in S_{b,z}$ exists. These
transitions are changed into $(s, q, q^{fresh}_{a,q})$, after that the
corresponding transitions of the form $(q^{fresh}_{a,q}, p, s^\prime)$ are added to
$\Gamma_{b,z}$.

\begin{figure}[htp]
\centering
\includegraphics[width=0.6\textwidth]{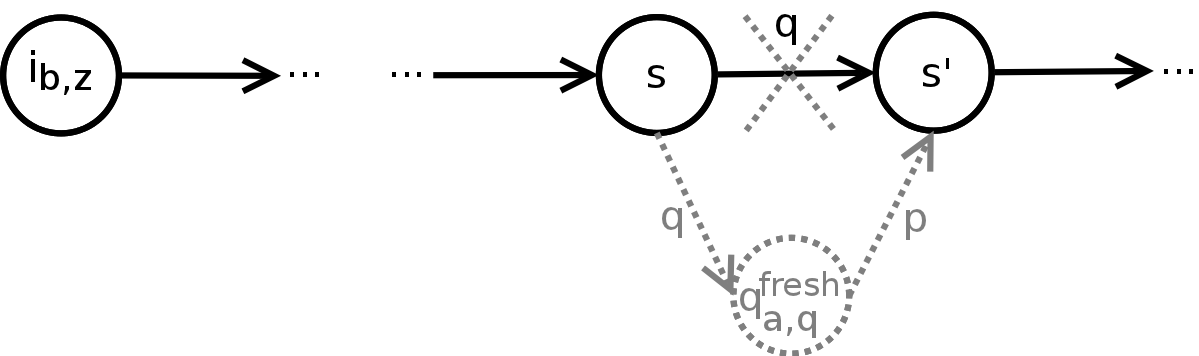}
\renewcommand{\figurename}{Figure}
\caption{The changes to the horizontal automaton due to rule $INS_{after}$ are
depicted as grey texts and dotted lines.}
\label{fig:INSafter_cap7.png}
\end{figure}

\subsubsection*{Replace: $RPL$}
For the rule $a(x) \rightarrow p \in R/A$ we need to modify each
horizontal language in which a state $q \in Q_L$ such that $L(B_{a,q}) \neq
\emptyset$ may occur. Each transition of the form $(s, q, s^\prime)$ included in
$\Gamma_{b,z}$, where $b \in \Sigma$ and $z \in Q_L$, is changed into $(s,
p, s^\prime)$.

\begin{figure}[htp]
\centering
\includegraphics[width=0.4\textwidth]{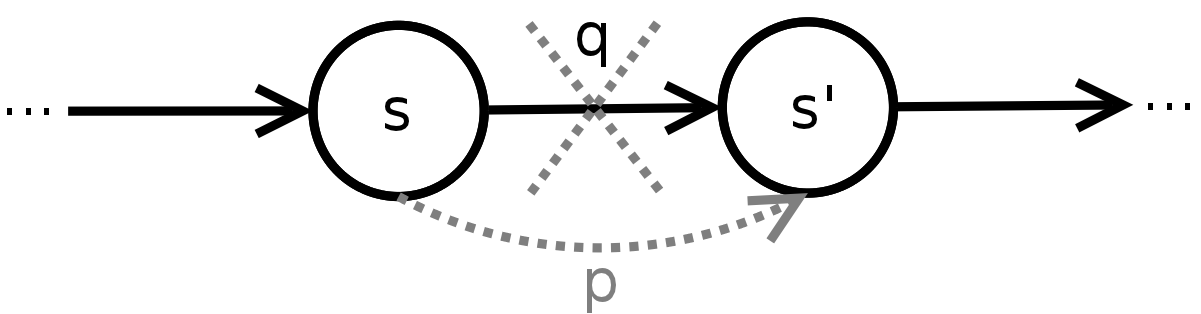}
\renewcommand{\figurename}{Figure}
\caption{The changes to the horizontal automaton due to rule $RPL$ are depicted
as grey texts and dotted lines.}
\label{fig:RPL_cap7.png}
\end{figure}

\subsubsection*{Delete: $DEL$}
For the rule $a(x) \rightarrow () \in R/A$ we need to modify every
horizontal language in which a state $q \in Q_L$ such that $L(B_{a,q}) \neq
\emptyset$ may occur. Each transition of the form $(s, q, s^\prime)$ in
$\Gamma_{b,z}$, where $b \in \Sigma$ and $z \in Q_L$, is changed into $(s,
\epsilon, s^\prime)$.

\begin{figure}[htp]
\centering
\includegraphics[width=0.4\textwidth]{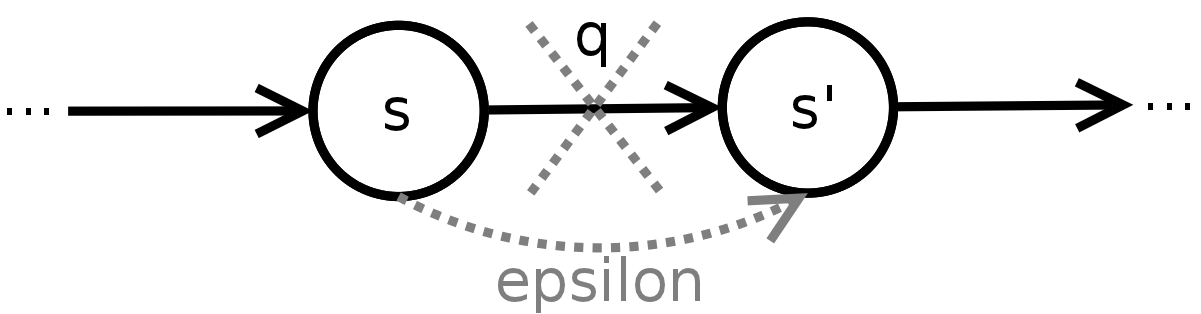}
\renewcommand{\figurename}{Figure}
\caption{The changes to the horizontal automaton due to rule $DEL$ are depicted
as grey texts and dotted lines.}
\label{fig:DEL_cap7.png}
\end{figure}

In the end, $\Delta^\prime$ is computed as 
$\Delta^\prime := \Theta \cup \{a(B_{a,q}) \rightarrow q \mid a \in \Sigma, q
\in
Q_L, L(B_{a,q}) \neq \emptyset)\}$.
The transitions of $\Theta$ ensures that $A^\prime$ is able to evaluate any
subtree belonging to $L$, the other transitions are used by $A^\prime$ for the
evaluation of the elements of $L(A)$ with the changes due to the update
operations. The test $L(B_{a,q}) \neq \emptyset$ excludes unnecessary
transitions.\\
\\
To preserve the tree structure of an XML document we need to avoid the
application of the operations $INS_{before}$, $INS_{after}$ and $DEL$, of the
form $a(x) \rightarrow pa(x)$, $a(x) \rightarrow a(x)p$ and $a(x) \rightarrow
()$, respectively, to any tree $t \in T(\Sigma)$ such that $t(\epsilon) = a$.

\subsubsection*{Insert into: $INS_{into}$} 
The simulation of the $INS_{into}$ rule requires some care. The rule
inserts a subtree in a nondeterministically chosen position in between the children
of a given node. Since the position is not known in advance we can only guess a
state $s$ of the horizontal automata and replace its outgoing transitions with
transitions passing through a fresh state. However we may need to consider an
automaton for every such state $s$. We describe next the $Post$ construction for
a given choice of $s$. The rule $a(xy) \rightarrow a(xpy) \in R/A$ leads to
change the automaton $B_{a,q}$, for each $q \in Q_L$ such that $L_{a,q} \neq
\emptyset$. A fresh state $q^{fresh}_{a,q}$ such that $q^{fresh}_{a,q} \notin
S_{a,q}$ is created and added to $S_{a,q}$. At this point, for each state $s \in
S_{a,q}$ reachable from $i_{a,q}$ through the transitions in $\Gamma_{a,q}$,
each transition of the form $(s, j, s^\prime)$ is changed into one of the form
$(s, j, q^{fresh}_{a,q})$, where $j \in P \cup Q_L$ and $s^\prime \in S_{a,q}$,
and transitions of the form $(q^{fresh}_{a,q}, p, s^\prime)$ are added to
$\Gamma_{a,q}$.

\begin{figure}[htp]
\centering
\includegraphics[width=0.6\textwidth]{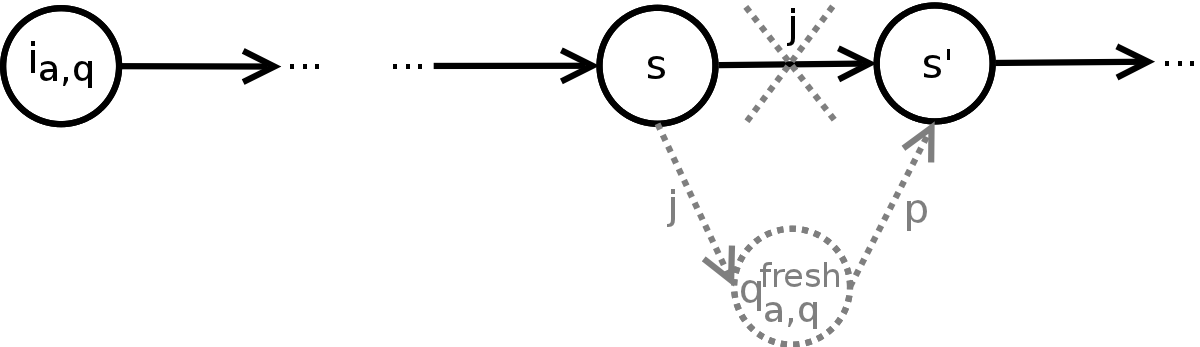}
\renewcommand{\figurename}{Figure}
\caption{The changes to the horizontal automaton due to rule $INS_{into}$ are
depicted as grey texts and dotted lines.}
\label{fig:INSinto_cap7.png}
\end{figure}

The need of guessing the right position in the horizontal automata in which 
inserting a fresh state generates several possible Hedge automata for each occurrence
of $INS_{into}$. However, in real implementations this operation often reduces either to 
$INS_{first}$ or $INS_{last}$. Thus in practical cases, this avoids the need of
introducing a search procedure in our HASA module.
\begin{es}
Suppose we have two NFHAs $A_L = (\Sigma_L, Q_L, Q^f_L, \Delta_L)$ and $A =
(\Sigma, P, P_f, \Theta)$ defined as follows:

\begin{itemize}
\item $\Sigma_L = \{a, b, c\}$ and $\Sigma = \{a, b, d\}$,
\item $Q_L = \{q_{a1}, q_{a2}, q_b, q_c\}$ and $P = \{g_a, g_b, g_d\}$,
\item $Q^f_L = \{q_{a1}, q_{a2}\}$ and $P_f = \{g_a\}$,
\item $\Delta_L = \{a({q_b}^*) \rightarrow q_{a2}, a({q_b}^*q_c) \rightarrow
q_{a1}, b(\epsilon) \rightarrow q_b, c(\epsilon) \rightarrow q_c\}$,
\item $\Theta = \{a({g_b}^+) \rightarrow g_a, b({g_b}^+|g_d) \rightarrow g_b,
d(\epsilon) \rightarrow g_d\}$.
\end{itemize}

The NFA used for the horizontal languages of the NFHA $A_L$ are:

\begin{itemize}
\item $B_{a,q_{a1}} = (Q_L, S_{a,q_{a1}} = \{p_b, p_c\}, p_b, \{p_c\},
\Gamma_{a,q_{a1}} = \{(p_b, q_b, p_b), (p_b, q_c, p_c)\})$,
\item $B_{a,q_{a2}} = (Q_L, S_{a,q_{a2}} = \{m_b\}, m_b, \{m_b\},
\Gamma_{a,q_{a2}} = \{(m_b, q_b, m_b)\})$,
\item $B_{b,q_b} = (Q_L, S_{b,q_b} = \{n\}, n, \{n\}, \Gamma_{b,q_b} = \{\})$,
\item $B_{c,q_c} = (Q_L, S_{c,q_c} = \{o\}, o, \{o\}, \Gamma_{c,q_c} = \{\})$.
\end{itemize}

It is clear that $L(A_L) = \{a(bc), a(bbc), \ldots, a(b\ldots bc), \ldots,
a, a(b), a(bb), \ldots, a(b\ldots b), \ldots\}$ and that $L(A)$ is the set of
the unranked tree where the root node is labelled with $a$, where the internal
nodes are labelled with $b$ and where the leaves are labelled with $d$.
Now we apply the update sequence $s = \{REN: b(x) \rightarrow a(x), INS_{first}: c(x)
\rightarrow c(g_ax), INS_{before}: c(x) \rightarrow g_ac(x)\}$ composed of
update operations of $R/A$ and we compute the NFHA $A^\prime = (\Sigma \cup
\Sigma_L, P \cup Q_L, Q^f_L, \Delta^\prime)$ such that $L(A^\prime) =
Post_{R/A}(L)$.

\begin{description}
\item[$REN: b(x) \rightarrow a(x)$;] the NFA $B_{a,q_b} = (P \cup Q_L$,
$S_{b,q_b} = \{n\}$, $n$, $\{n\}$, $\Gamma_{b,q_b} = \{\})$ is defined and
all the occurrences of label $b$ in the horizontal rules are replaced
with label $a$.

\item[$INS_{first}: c(x) \rightarrow c(g_ax)$;] the NFA $B_{c,q_c}$ is changed
into $(P \cup Q_L, \{q^{fresh}_{c,q_c}, o\}$, $q^{fresh}_{c,q_c}$, $\{o\}$,
$\{(q^{fresh}_{c,q_c}, g_a, o)\})$.

\item[$INS_{before}: c(x) \rightarrow g_ac(x)$;] the NFA $B_{a,q_{a1}}$ is
changed into $(P \cup Q_L, S_{a,q_{a1}} = \{p_b, p_c, q^{fresh}_{c,q_c}\}, p_b,
\{p_c\}$,\\ $\Gamma_{a,q_{a1}} = \{(p_b, g_a, q^{fresh}_{c,q_c}),
(q^{fresh}_{c,q_c}, q_c, p_c), (p_b, q_b, p_b)\})$.
\end{description}

\begin{figure}[htp!]
\centering
\subfigure[The parallel application of $REN$, $INS_{first}$ and $INS_{before}$
changing $t \in L$ into $t^\prime \in L(A^\prime)$.]{%
  \label{fig:es1_cap7}
  \includegraphics[width=0.55\textwidth]{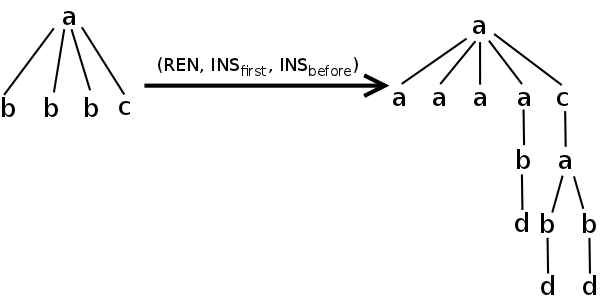}
}
\subfigure[The accepting computation of the NFHA $A^\prime$ over $t^\prime$.]{%
  \label{fig:es2_cap7}
  \includegraphics[scale=0.30]{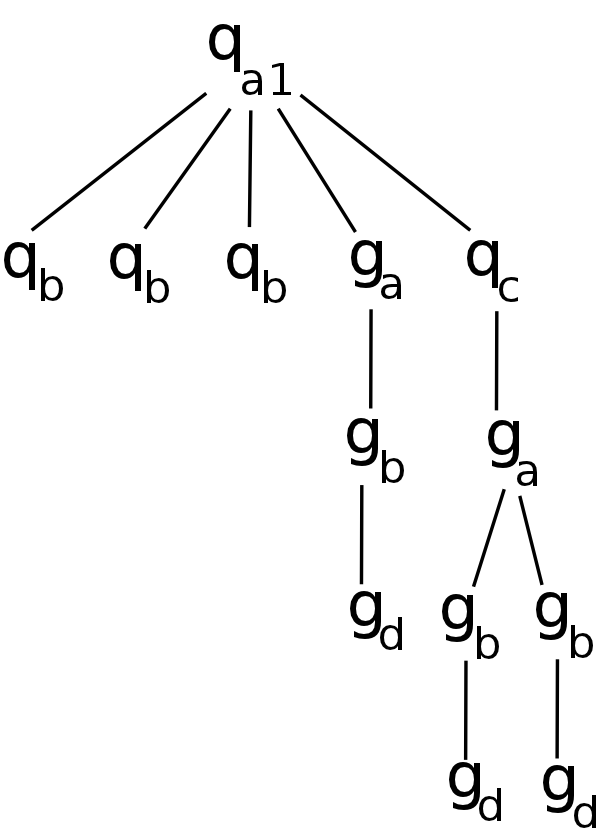}
}%
\caption{An example of update from tree $t$ to $t^\prime$ (left) and the
accepting computation of the automaton $A^\prime$ accepting the updated language
over updated tree $t^\prime$ (right).}
\end{figure}

In Figure \ref{fig:es1_cap7} we can see an example of application of the update
operations $REN$, $INS_{first}$ and $INS_{before}$ that transform tree $t \in L$
into $t^\prime \in L(A^\prime)$. In Figure \ref{fig:es2_cap7} we can see an
accepting computation of the NFHA $A^\prime$ related to tree $t^\prime$. $\Box$

\end{es}

\section{Related Work and Conclusions}
\label{sec:conclusionrw}
We have developed a Java prototype based on the LETHAL Library\footnote{LETHAL
is available at \url{http://lethal.sourceforge.net/}}.
The experiments started from the XML benchmark used in the \emph{XMark
Benchmark} Project\footnote{The benchmark and related schema are available at
\url{http://www.xml-benchmark.org/}}. We tested the complete set of update
primitives, both in isolation and in a sequence of updates. The modified schema,
that is intended to be obtained from a schema update sequence, is manually
generated. A valid (resp. invalid) sequence of document updates is tested by
means of our symbolic computation and by means of inclusion test for HA provided
by the library. The $Post$ algorithm works on a representation of horizontal
languages as regular expressions (we adapted our algorithm to deal with it) and
then computes a new HA. This is due to limitations of the LETHAL library, which
is not designed for low-level manipulations of automata but only for the
application of common HA operations (inclusion, union, intersection, etc.).
Despite inclusion test complexity for NFHA is ExpTime-Complete \cite{tata2007},
the execution times of the $Post$ computation and of the inclusion test on the
considered XML benchmark are negligible (less than 1s) even with a na\"{\i}ve
implementation. These results are not surprising because the automaton size
depends on the corresponding schema  size, that is usually limited (in
terms of labels and productions) in practical schemas. In addition, schema
size in not comparable with the one of the associated document collection (in
terms of document number and size). The results show the potential of our
proposal for a practical usage as a support for static analysis of XML updates.
Before addressing possible extension, we discuss next some related work.

Concerning related work on static analysis, the main formalization of schema
updates is represented by \cite{benedikt_cheney}, where the authors take into
account a subset of XQUF which deals  with structural conditions imposed by tags
only. Type inference for XQUF, without approximations, is not always possible.
This follows from the fact that modifications that can be produced using this
language can lead to nonregular schemas, that cannot be captured with existing
schema languages for XML. This is the reason why \cite{benedikt_cheney}, as well
as \cite{Touili_computingtransitive}, computes an over-approximation of the type
set resulting from the updates. In our work, on the contrary, to produce an
exact computation we were forced to cover a smaller subset of XQuery Update
features: \cite{benedikt_cheney}, indeed, allows the use of XPath axes to query
and select nodes, allowing selectivity conditions to be mixed with positional
constraints in the request that a given pattern must satisfy. In our work, as well as
in \cite{Touili_computingtransitive} and \cite{jacqemard_rusinowitch},
we have considered update primitives only, thus excluding complex expressions
such as ``for loops'' and ``if statements'', based on the result of a query.
These expressions, anyway, can be translated into a sequence of primitive
operations: an expression using a ``for loop'', for instance, repeats $n$ times
a certain primitive operation, and therefore can be simulated with a sequence of
$n$ instances of that single primitive operation.\footnote{The interested reader
could refer to \cite{benedikt_cheney} (Section ``Semantics''), where a
translation of XQUF update expressions into a pending update list, made only of
primitive operations, is provided, according to the W3C specification
\cite{xqueryupdatefacility_w3c}.} However, tests for loops and conditional
statements based on query results over documents are of course not expressible
working only at schema level.
\textit{Macro Tree Transducers} (MTT)
\cite{ManethBPS05} can also be applied
to model XML updates as in the \textit{Transformation Language} (TL), 
based on \textit{Monadic Second-Order logic} (MSO).
TL does not only generalize XPath, XQuery and XSLT, but can also be simulated
using macro tree transducers. The composition of MTT and their property of
preserving recognizability for the calculation of their inverses are exploited
to perform \textit{inverse type inference}: they pre-compute in this way the
pre-image of ill-formed output and perform type checking simply testing  whether the
input type has some intersection with the pre-image.\footnote{Note that tree
languages are closed under intersection and that the emptiness test is decidable
for them.} 
Their system, as ours, is exact and does not approximate the
computation, but, in contrast to our method there is a potential implementation
problem (i.e., an exponential blow-up) for the translation of MSO patterns into
equivalent finite automata, on top on which most of their system is developed,
even if MSO is not the only suitable pattern language that can be used with
their system. Thus, our more specific approach, focused on a specific set of
transformations, allows for a simpler (and more efficient) implementation.

Our approach complements work on XML schema evolution developed in the
XML Schema context \cite{xsym07,icde11}, where validity preserving schema updates
are identified and automatic adaptations identified, when possible. In case 
no automatic adaptation can be identified, the use of user-defined adaptation
is proposed, but then a run-time (incremental) revalidation of all the adapted 
documents is needed. 
Similarly, in \cite{genevestoit2012} a unifying framework for determining the
effects of XML Schema evolution both
on the validity of documents and on queries is proposed. The proposed system
analyzes various scenarios in which forward/backward compatibility of schemas is
broken. In \cite{RaghavachariS07} a related but different problem is
addressed: how to exploit the knowledge that a given document is valid with
respect to a schema $S$ to (efficiently) assess its validity with respect to a
different schema $S^\prime$.
Finally, document update transformation is addressed in \cite{BonevaCGRTS11}, 
which investigates how to rewrite (document) updates specified on a view
into (document) updates on the source documents, given the XML view definition.

The present work can be extended along several directions.
Node selection constraints for update operations could be refined, for
example using \textit{XPath} axes \cite{xpath_w3c} and the other features
offered by XQUF. It may be interesting to integrate the existing Java
prototype of the framework with XML schema evolution tools like \textit{EXup} 
\cite{icde11}.
Moreover, when the schema update operation sequence is known,
a heuristic to automatically extract a sequence of update operations 
that will ensure document validity
with respect to the new schema, relieving the user from specifying the
appropriate sequence, and  generalizing the  automatic adaptation approach
currently supported in EXup, could be devised.
Finally, support for commutative trees, in which the order of the children
of a node is irrelevant, could be added. This feature would allow the
formalization of the
\textit{all} and \textit{interleave} constructs of XML Schema
\cite{xmlschema_w3c} and Relax NG \cite{relax_ng_spec,relax_ng_vandervlist},
respectively, and the overcome of the need of considering several alternative automata
for the $INS_{into}$ operation. Sheaves Automata, introduced in \cite{sheaves_automata}, are able
to recognize commutative trees and have an expressiveness strictly greater than
the HA considered in this work. The applicability of these automata in our
framework needs to be investigated.

\bibliographystyle{eptcs}
\bibliography{thesis}


\end{document}